\documentclass[journal]{IEEEtran}
\usepackage[pdftex]{graphicx}
\usepackage{caption}
\usepackage[font={footnotesize}]{caption}
\usepackage[table]{xcolor}
\usepackage{array,dcolumn,booktabs}
\ifCLASSINFOpdf
\else
\fi
\usepackage[cmex10]{amsmath}

\usepackage{setspace}
\usepackage[linesnumbered,ruled]{algorithm2e}

\makeatletter
\def\footnoterule{\kern-3\p@
  \hrule \@width 2in \kern 2.6\p@} 
\makeatother

\begin{document}
%
\title{GlidarCo: gait recognition by 3D skeleton estimation and biometric feature correction of flash lidar data}
%
%
%

\author{Nasrin~Sadeghzadehyazdi,
Tamal~Batabyal, Nibir K. Dhar,  
B.~O.~Familoni,
K.~M.~Iftekharuddin,
Scott~T.~Acton
}
\maketitle

\begin{abstract}
Gait recognition using noninvasively acquired data has been attracting an increasing interest in the last decade. Among various modalities of data sources, it is experimentally found that the data involving skeletal representation are amenable for reliable feature compaction and fast processing. Model-based gait recognition methods that exploit features from a fitted model, like skeleton, are recognized for their view and scale-invariant properties. We propose a model-based gait recognition method, using sequences recorded by a single flash lidar. Existing state-of-the-art model-based approaches that exploit features from high quality skeletal data collected by Kinect and Mocap are limited to controlled laboratory environments. The performance of conventional research efforts is negatively affected by poor data quality. We address the problem of gait recognition under challenging scenarios, such as lower quality  and noisy imaging process of lidar, that degrades the performance of state-of-the-art skeleton-based systems. We present GlidarCo to attain high accuracy on gait recognition under the described conditions. A filtering mechanism corrects faulty skeleton joint measurements, and robust statistics are integrated to conventional feature moments to encode the dynamic of the motion. As a comparison, length-based and vector-based features extracted from the noisy skeletons are investigated for outlier removal. Experimental results illustrate the efficacy of the proposed methodology in improving gait recognition given noisy low resolution lidar data.
\end{abstract}

{\let\thefootnote\relax\footnote{{Nasrin Sadeghzadehyazdi, Tamal Batabyal, and Scott T. Acton are with the Charles L. Brown Department  of  Electrical and Computer Engineering, University of Virginia, Charlottesville, VA 22904-4743, USA  (e-mails: ns8va@virginia.edu, tb2ea@virginia.edu and acton@virginia.edu.)

Nibir K. Dhar, and B. O. Familoni are with Night Vision and Electronic Sensors Directorate Fort Belvoir, USA

K. M. Iftekharuddin is with department of electrical and computer engineering, Old Dominion University, 231 Kaufman Hall Norfolk, VA 23529, USA (e-mail: kiftekha@odu.edu.)

Distribution Statement A: Approved for public release. Distribution is unlimited

IEEE  Copyright  Notice: {\scriptsize\textcircled{c}}IEEE  2019  Personal  use  of  this  material  is permitted. Permission from IEEE must be obtained for all other uses, in any current or future media,  including reprinting/republishing this material for advertising or promotional purposes, creating new collective works, for resale or redistribution to servers or lists, or reuse of any copyrighted component of this work in other works.}}}


%
\IEEEpeerreviewmaketitle

\section{Introduction}
%
%
%
%

\IEEEPARstart{G}{ait} identification has received an increasing interest in the last decade due to the various applications in areas ranging from intelligent security surveillance and identifying person of interest in criminal cases, to designated smart environments \cite{jain2006biometrics,boulgouris2005gait}. Besides, gait analysis plays an important role to quantify the severity of certain motion-related diseases like Parkinson \cite{del2016validation}. Gait recognition aims to tackle the identification problem based on the way people walk and early studies findings in the medical and psychology have shown the uniqueness of gait to individuals \cite{cutting1977recognizing, charalambous2014walking}. While the iris \cite{daugman2009iris}, face \cite{turk1991face, schroff2015facenet}, and fingerprint \cite{maltoni2009handbook} provide some of the most efficacious biometrics for person identification with high recognition accuracy, they require the cooperation of subjects as well as availability of high quality data. In real life however, there are many scenarios in which the subjects cannot be controlled, there is no contact between subjects and sensors, or access to the high quality data is not possible. Under such circumstances, biometrics that can be extracted from the gait have shown promising results in several studies \cite{lee2002gait,preis2012gait,sinha2013person}. Features extracted from gait are resilient to changes in clothing or lighting conditions compared to color or texture which are among prevalent features for person identification. While patterns of walking may not be necessarily unique to individuals in practice, a combination of biometric-based static attributes along with the motion analysis of certain joints can create an effective set of features to recognize one individual from another. 

\begin{figure}
[!t]
\centering
	\includegraphics[width=\linewidth]{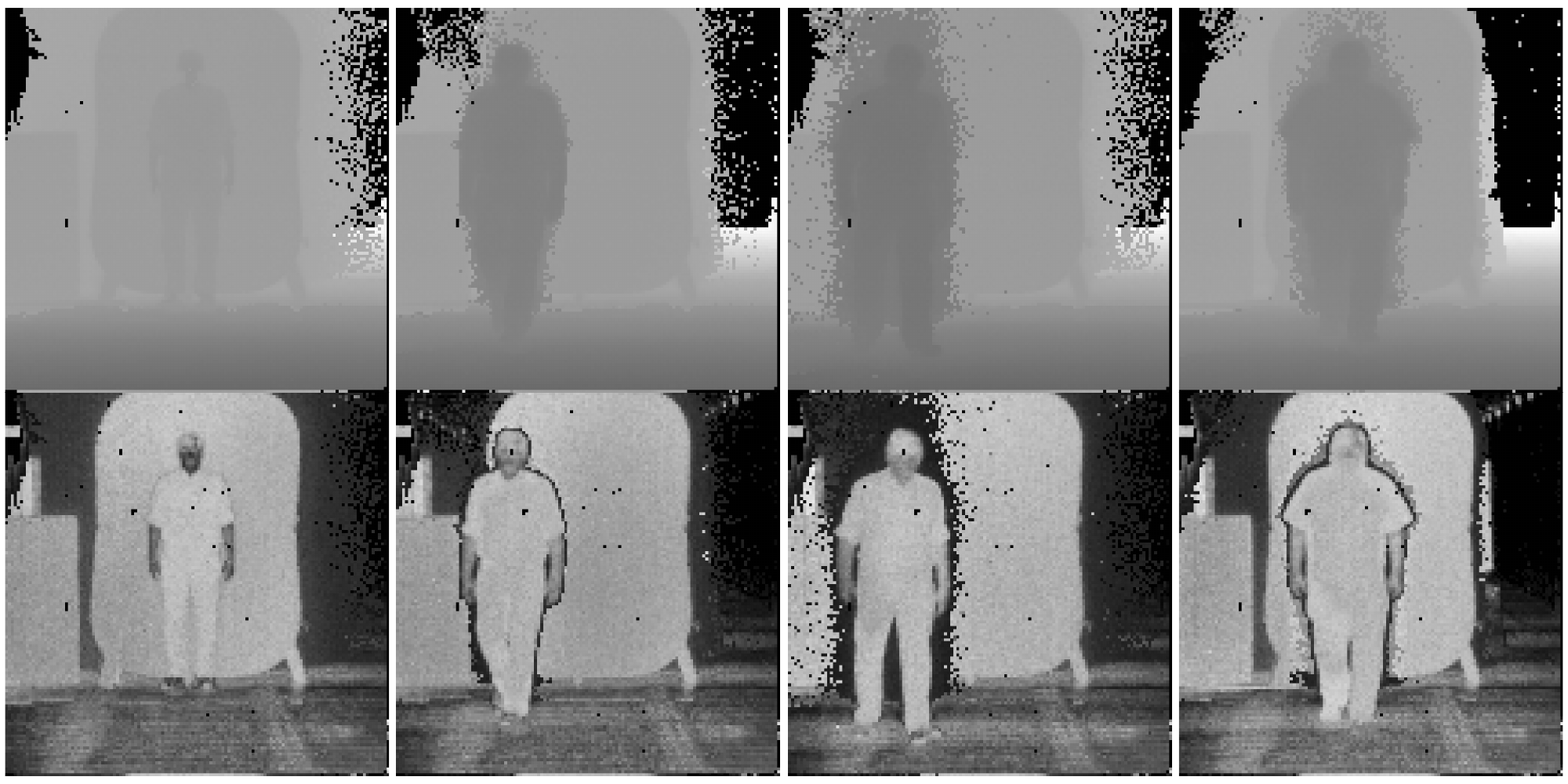}
	\caption[Sample frames of lidar]{Sample frames of lidar data. The top and bottom rows show range and intensity data, respectively.}
    \label{fig:SampleFramesOFLidar}
\end{figure}

In recent years, depth cameras have become popular for gait analysis mainly due to their ability to provide a three dimensional depiction of the scene \cite{batabyal2016ugrad,batabyal2015ugrasp,clark2013concurrent}. Unlike their optical counterparts, depth cameras like lidar and Kinect can provide depth information that is not sensitive to illumination and changing in the lighting conditions that are among major issues in uncontrolled environments. This makes depth camera an ideal candidate for long-term person identification that rely on features such as time-invariant biometrics. In this work, we utilize the flash lidar technology to collect data. A flash lidar camera uses pulsed laser to illuminate the whole scene and simultaneously record range (depth) and intensity information. Figure \ref{fig:SampleFramesOFLidar} shows sample frames of the collected intensity and range data by the flash lidar camera. Since the laser beams can be focused to suit the objects of interest, a flash lidar camera can provide detailed depth imaging of the scene. This property of flash lidar has lead to extensive applications in areas such as autonomous vehicles, atmospheric physics, archaeology, forestry, geology, geography, seismology, space missions, and transportation. 

Video-based gait recognition approaches are generally divided into two main categories, model-based, and model-free methods. Model-free methods rely on features that can be obtained from clean silhouettes \cite{kale2004identification,benedek20143d}. They are easy to implement and computationally less expensive compared to their model-based counterparts. However, model-free methods are not view and scale invariant, and require recordings from multiple angles that is not always feasible from applicability point of view. Using skeleton for gait recognition is categorized as a model-based approach for person identification. Most of the existing methods rely on fitting a model, usually a skeleton, to human silhouettes \cite{fujiyoshi2004real,bobick2001gait}. The main issue with the model-based methods is the fact that in general, model fitting is a computationally expensive process. While such difficulties are not an issue with structured lighting approaches such as Kinect due to the direct estimation of joints coordinates, the working range of Kinect is limited. Furthermore, the range information of Kinect is not reliable in outdoor environments, because it is not easy to differentiate the infrared light of the sensor from the high intensity infrared of environment \cite{fankhauser2015kinect,zennaro2014evaluation}. To curb the computational complexity of model-based methods, several studies rely on high-quality real-time skeleton joints data generated by Mocap \cite{krzeszowski2014dtw, balazia2017human}. However, in terms of applicability, Mocap is limited to a laboratory environment which is a major drawback. Unlike Mocap, flash lidar has been extensively used for outdoor applications. Compared with Kinect, a flash lidar camera has a drastically extended range ($>1000$~meters) and its performance is not degraded in outdoor environments due to the high irradiance power of pulsed laser generated by lidar compared with the background \cite{horaud2016overview}. 

With a limited number of studies, the only existing lidar-based person identification works in the literature are model-free and rely on background subtraction to extract human silhouette from the point cloud data provided by Velodyne’s Rotating Multi-Beam
(RMB) lidar system \cite{benedek2018lidar,galai2015feature,benedek20143d}. In this work however, we take a model-based approach, leveraging OpenPose, a pre-trained deep network \cite{cao2016realtime} to extract a skeleton model from the intensity information. Using camera properties and the depth data, the provided skeleton joint coordinates are transferred into real-world coordinates. The work presented here can be employed for gait analysis in different applications; for instance to improve the classification results by including gait information in our previous work \cite{sadeghzadehyazdi2016graph}.

This shift of the modality from the structured (image/video) to unstructured (skeleton) data type provides benefits in terms of data compaction, computation, storage, scalability, and recognition accuracy. Furthermore, the skeleton-related attributes mimic actual physical traits in human body and can be utilized as a soft biometric ID for the individuals. Our visual system does not extricate details of clothing texture, or the skin tone of the person who is walking. Rather, it focuses on certain body parts (joints, limbs), and tries to reconstruct the anatomy and locomotion. Such biological cues are exploited in  ~\cite{kovashka2010learning,ofli2014sequence}. In addition, there has been a surge of studies to find a suitable model for such purpose~\cite{vemulapalli2014human, batabyal2015action, evangelidis2014skeletal}.

Existing successful model-based methods take advantage of high-quality skeleton data  provided by Kinect or Mocap and avoid the challenge of erroneous features. However, as we mentioned earlier, these modalities are not a proper choice for real-world applications. In contrast with Kinect and Mocap, the data collected by a flash lidar camera is noisy and has low resolution that degrades the performance of skeleton extraction systems. Features that are computed from the faulty skeleton models are plagued with erroneous measurements that in turn present a major challenge for a successful gait recognition. In this paper, our main goal is to answer the following question, "When the collected data is noisy to a level that a considerable number of fitted skeleton models contain missing or erroneous joints, is it still possible to identify gaits with a high accuracy and precision?". In particular under the described condition, "Can we avoid the common approach of removing noisy data, and correct the faulty skeletons, instead?" To address these questions, we present GlidarCo, a methodology to correct for the faulty and missing measurements of joint coordinates, and integrate the robust statistics to improve gait recognition using the noisy, low resolution flash lidar data. 

Our contributions are fourfold. First, we present a model-based approach for gait recognition using flash lidar data that is close to real-time. Second, we present a filtering mechanism that exploits robust statistics and shape-preserving interpolation to correct for faulty and missing measurements of joint coordinates. Third, we integrate robust statistics with the traditional feature moments to incorporate the motion dynamics over the gait cycles. Fourth, as an alternative method for applications where data elimination is not an issue, we investigate features extracted from noisy skeletons for outliers, and present a modification of the Tukey's method for vector-based feature vectors. The latter contribution is an effort to follow the traditional practice of removing noisy data and perform classification on the remaining clean data. In particular, we aim to compare the results from outlier removal method, with an unorthodox effort that seeks to correct the erroneous data. We must emphasize the importance of the latter, as it preserves the original data, that is costly to collect in many applications. An extensive experimental investigation demonstrates the efficacy of the proposed methodology in improving the performance of both length-based and vector-based features for gait identification using the flash lidar data. 

The rest of this paper is presented as follows. In the next section, we will outline the related work. Next, we will describe the proposed methodology in detail in the Methods section. The results and discussion describes a thorough experimental investigation into the efficacy of the proposed method and compares the performance of multiple set of features, including state-of-the-are methods in the context of gait recognition before and after data correction. Finally, we summarize in the conclusion section.

\section{related work}
Model-based methods fit a model, like a skeleton, to human body and use the features extracted from the fitted model to identify gaits. Model fitting is generally a complex and computationally expensive process. To avoid such difficulties, many studies leverage Kinect as a marker-less motion capture tool, that generates a real-time high quality intensity and depth data, along with joint information of skeleton. In general, these methods are based on identifying gait cycle and calculating static anthropometric-based features like bone lengths and height, gait features like step length and gait cycle or angle between selected body joints over each gait cycle. Statistics like mean, maximum, and standard deviation of the collected attributes are computed over each gait cycle and utilized as feature identifiers.  

Using maximum, mean, and standard deviation of a set of lower body angles over a half gait cycle as features and K-Means clustering algorithm, Ball \textit{et al}. \cite{ball2012unsupervised} acquired an accuracy of 43.6\% on a dataset collected from four subjects. In \cite{preis2012gait}, authors used a set of static features plus two gait features and achieved an accuracy of 90\% on a dataset collected from nine subjects walking from right to left in front of a Kinect camera. Araujo \textit{et al}. \cite{araujo2013towards} used eleven static anthropometric features and investigated the effect of different subset of features in gait recognition. They also compared the performance of four different classifiers on a dataset collected from eight different subjects. An average accuracy of 98\% was obtained only when the training and test samples contained the same type of walking pattern. Sinha \textit{et al}. \cite{sinha2013person} proposed a set of area-based features plus distance between different body segment centroids and combined these attributes with features in \cite{ball2012unsupervised} and \cite{preis2012gait} and obtained a higher accuracy compared with the work of Ball and Preis on a dataset of ten subjects. Kumar and Babu \cite{kumar2012human} proposed a set of covariance-based measures on the trajectory of skeleton joints, and acquired an accuracy of 90\% on a dataset of 20 subjects. Dikovski \cite{dikovski2014evaluation} evaluates the performance of different features like angles of lower body joints, distance between adjacent joints, height and step length over one gait cycle. Relative distance and relative angles are computed between selected body joints and compared together utilizing the Dynamic Time Warping (DTW) algorithm by Ahmed \textit{et al}. \cite{ahmed2015dtw}. Ali \textit{et al}. \cite{ali2016applying} compute triangles formed by lower body joints during motion and utilize mean of the areas during one gait cycle. In \cite{yang2016relative} Yang \textit{et al}. use a set of anthropometric and relative distance-based features for identification. 
\begin{figure}[!t]
\centering
\includegraphics[width=\linewidth]{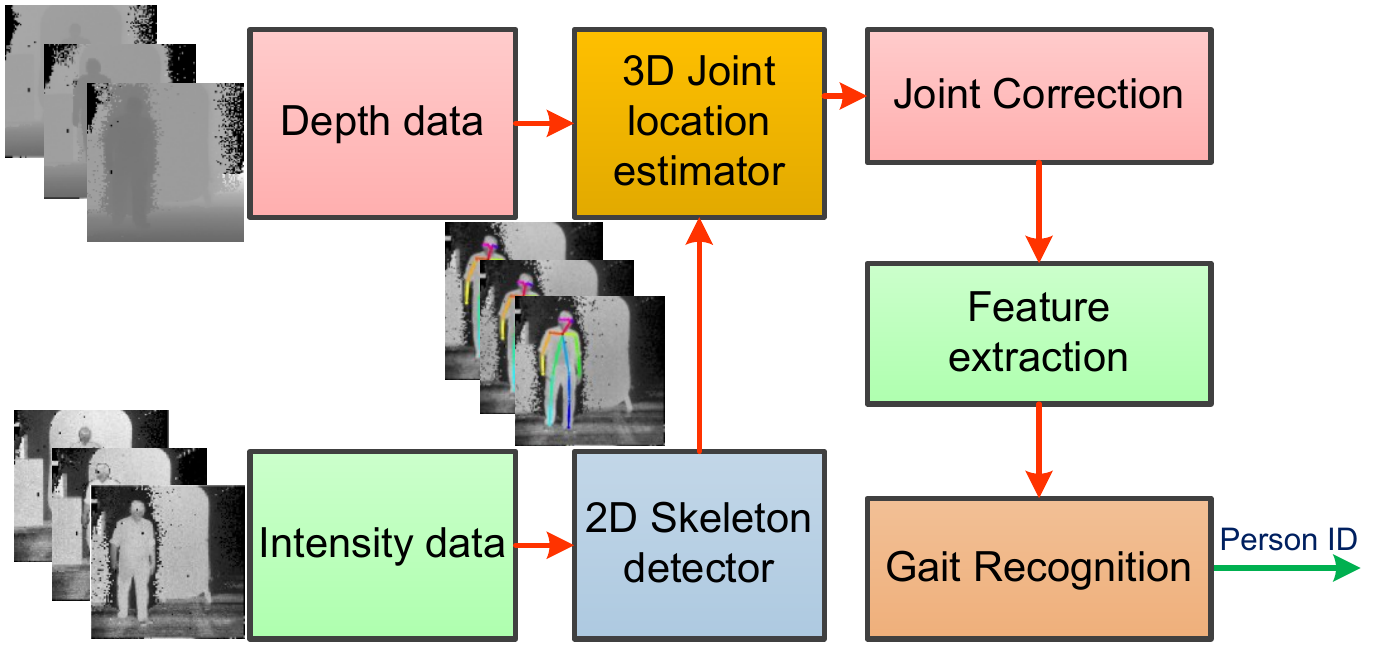}
\makeatletter
\caption[Joint correction flowchart] {Pipeline for gait recognition using joint correction criterion of GlidarCo}   
\label{fig:flowChart}
\end{figure}
\begin{figure}[!t]
\centering
\includegraphics[width=\linewidth]{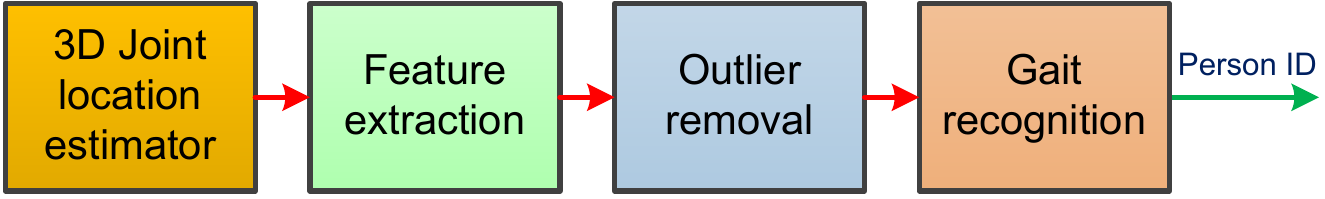}
\makeatletter
\caption[Outlier removal] {Pipeline for outlier removal. Inputs to \textit{"3D Joint location estimator"} remain the same as in Figure \ref{fig:flowChart} } 
\label{fig:outlierRemoval}
\end{figure}

The majority of previous model-based studies exploit sequences that were recorded on a limited walking patterns. Subjects walk on straight lines and training and test sequences include the same patterns of walking. In this work, we consider a case that involves different patterns of walking. In particular, we select disparate walking patterns for training and test sequences. 

This paper is an extension of our previous work \cite{sadeghzadehyazdi2019glidar3dj}, with an improvement on joint coordinate filtering and per-frame identification. Furthermore, we propose a new method to integrate the dynamic of the motion. We also present an outlier removal method for vector-based feature vectors that can be employed in applications that data removal is not an issue. In addition, we evaluate the proposed methodologies on two different feature vectors, and compare with more state-of-the-art relevant methods.
\begin{figure}
[!b]
\centering
	\includegraphics[width=\linewidth]{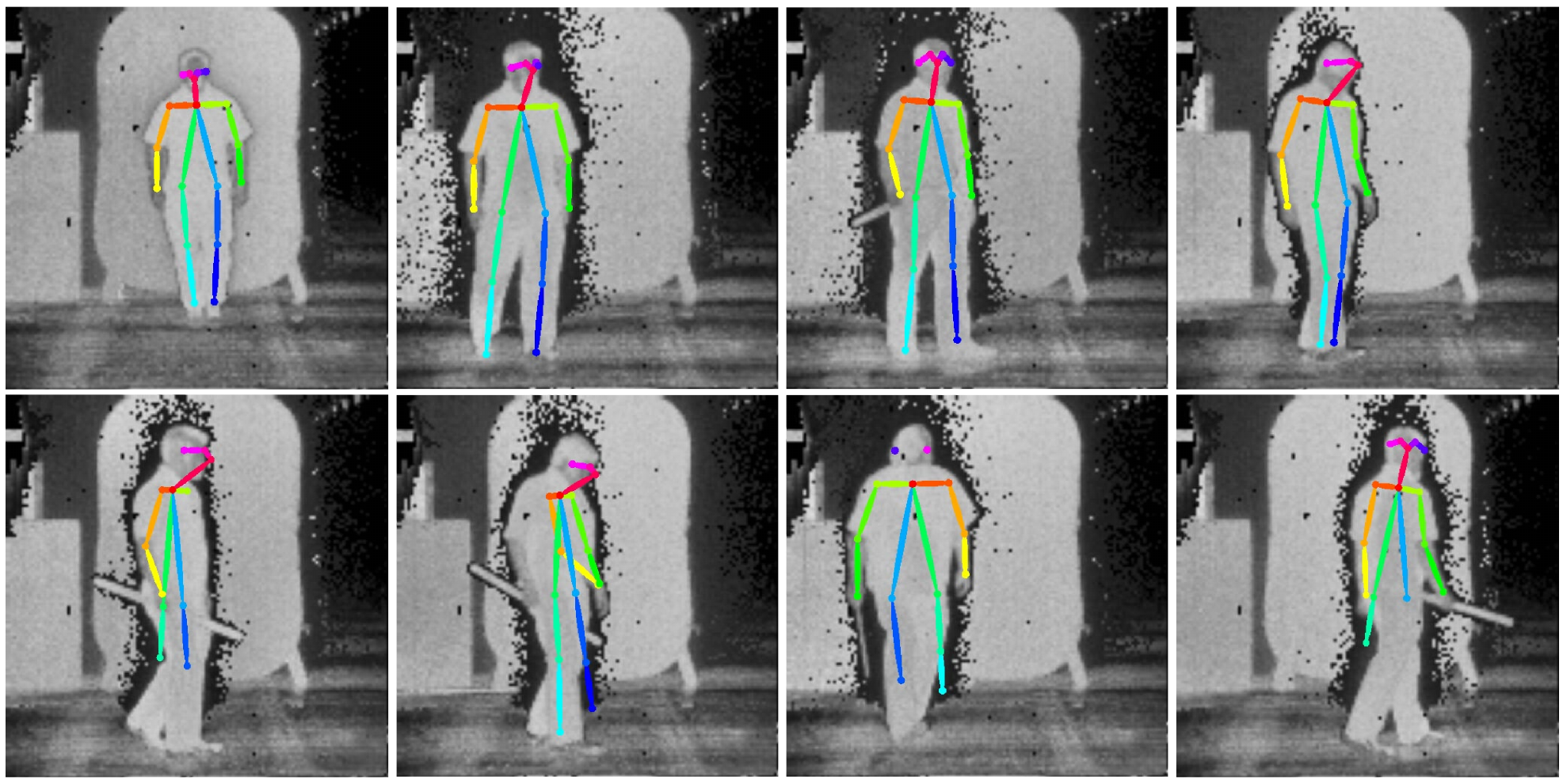}
	\caption[Sample frames with Skeleton]{Top row: sample frames with correctly detected skeletons, bottom row: frames with faulty skeletons}
    \label{fig:FramesWithSkeleton}
\end{figure}

In our dataset, recorded by a single flash lidar camera, several factors diminish the quality of features that are computed from the resulting joints. As the subjects proceed toward the camera, range data are affected by noise. The lack of color in the intensity data, and similarity between human clothing, background and skin are some of the other elements that can negatively affect the quality of detected poses and consequently the feature vectors. A common approach in the existing studies involves the removal of outlier noisy data that are generated as a result of faulty measurements. Further processing is applied on the remaining higher quality collection of joint data. In this paper, we propose an automated outlier removal procedure. However, with the shortage of data being a major challenge in many real-world surveillance scenarios, data removal will only exacerbate the data scarcity problem. In other words, while outlier removal can be a proper solution to gather higher quality data to begin with, it is not the best choice when data elimination can raise issues. We aim to address this problem by proposing a filtering mechanism that corrects erroneous joint data, instead of eradicating them. Figure \ref{fig:flowChart} and \ref{fig:outlierRemoval} present the pipeline of joint correction and outlier removal methodologies, respectively.   

To prove the efficacy of the proposed methodology, we will compare the performance of two different sets of features, length-based and vector-based features, and four state-of-the-art works, before and after joint correction. Furthermore, as an alternative for applications in which data elimination is not an issue, we also consider automatic outlier removal and compare it with the proposed joint correction on improving the gait identification accuracy.  
\section{Methods}

\subsection{Overview of method}
\begin{figure}[!t]
\centering
\includegraphics[width=.8\linewidth]{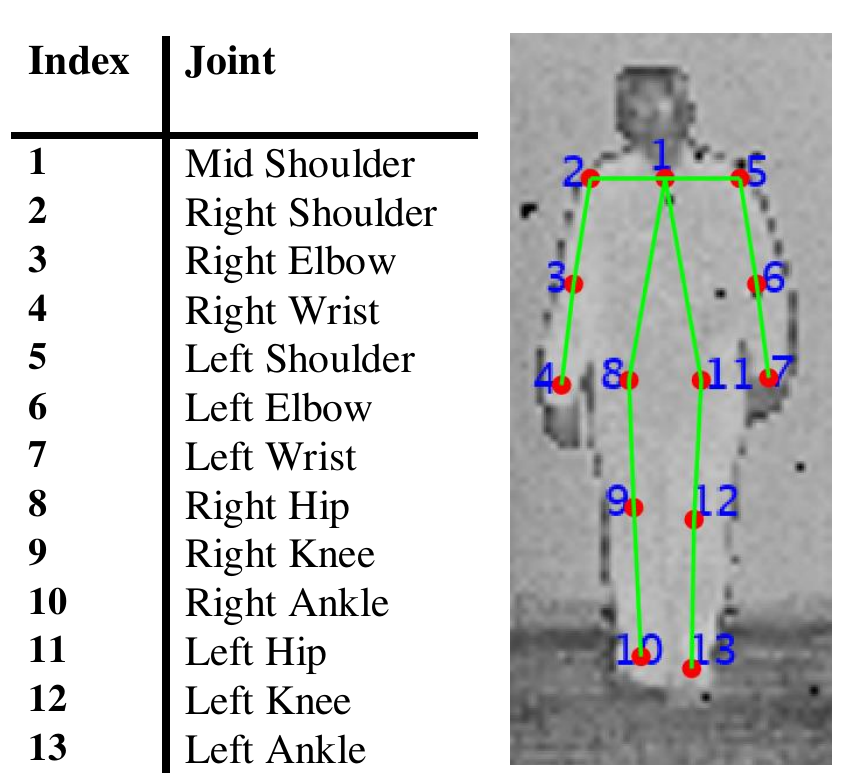}
\makeatletter
\caption[Joint and skeleton] {The skeleton model we use in this work. Left: index of each joint in the skeleton model. Right: skeleton model in a sample frame.} 
\label{fig:JointAndSkeleton}
\end{figure}

Figure \ref{fig:flowChart} describes the workflow of the proposed gait recognition methodology using flash lidar data. For a lidar sequence $V$ with $f$ frames, there exists $I = [I_{1}, I_{2}, ..., I_{f}]$, and $R = [R_{1}, R_{2}, ..., R_{f}]$, where $I_{i}$ and $R_{i}$ represent intensity and range data at frame $i$. Images are preprocessed to reduce noise and are fed into a 2D skeleton detector. We leverage OpenPose a state-of-the-art real-time pose detector to fit a skeleton model and extract the location of body joints. In Figure \ref{fig:FramesWithSkeleton}, the top row shows examples of correctly detected skeleton joints. As we can see in this figure, OpenPose provides a skeleton model of 18 joints, where 5 of the joints represent nose, eyes, and ears. However, the model that we adopt in this paper only considers 13 joints. The reason for such choice is the fact that face joints are missing from a large majority of our samples. Furthermore, the facial joints do not convey useful information for gait recognition. Figure \ref{fig:JointAndSkeleton} illustrates the skeleton model that we use in this work. Given $I_{i}$ as the input to the skeleton detector, the output is the joint location coordinates that can be represented with the following vectorized form
\begin{equation}
\label{eq:OpenPoseOutput}
J_{i}=[x_{k},y_{k}]^{M_{j}}_{k=1}\in \Re^{2N} 
\end{equation}
where $(x_{k},y_{k})$ are the coordinates of the $kth$ joint in the image frame of reference, and $M_{j}$ represents the number of joints. Using the range data and the properties of the lidar camera, we can project the 2-dimensional coordinates of joints into real-world coordinates. $L^{i}_{j}$, the real-world location of joint $i$ in the direction $j$ can be calculated according to the following equation
\begin{equation}
\label{eq:realWorld}
L^{i}_{j}  =  \frac{2}{N_{pixels}}\times\tan(\frac{\theta_{aov}}{2})\times Lp^{i}_{j} \times D^{i}_{camera}
\end{equation}
where $N_{pixels}$ is the number of pixels in the $j$ direction, $\theta_{aov}$ represents the angle of view, and $D^{i}_{camera}$ is the range value of joint $i$. $Lp^{i}_{j}$ represents the location of joint $i$ in the direction $j$ in the image coordinate system. Here $j$ is in the $x$ or $y$ direction, and the $L^{i}$ in the $z$ direction equals to the depth value at the location of the joint $i$.

As we discussed earlier, the quality of the resulting skeleton and the joint localization are negatively affected by several factors. The features that are computed using the acquired skeletons are plagued with erroneous measurements. Therefore, gait recognition based on the computed faulty skeletons results into a high rate of false positives. To resolve this problem, we present a filtering mechanism that employs robust statistics and shape-preserving interpolation to correct for faulty measurements in time sequences of joint coordinates values. This filter will improve the quality of the joint localization and ultimately enhance the gait recognition accuracy. As an alternative approach for the joint location correction, we employ the Tukey method to detect and remove length-based and vector-based feature vectors. In particular, we present a modification for vector-based outlier detection using the Tukey method. The following subsection gives the description of the filtering mechanism, which is followed by the outlier removal subsection.
\subsection{Filtering of joint location}
Let $L$ be a matrix of the size of $39\times F_{n}$, where each row represents the time sequence of one joint in one of the direction of $x,y,$ and $z$, extended over $F_{n}$ frames. Since each skeleton consists of $13$ joints, there are in total $39$ joint coordinate time sequences. In order to correct for missing joint location values and noisy outliers in a given video, we perform filtering of joint location on each row of the corresponding $L$ matrix. Let $L_{m}$ represent the $m$\textit{-th} row of $L$
\begin{equation}
\label{eq:JointLocation}
L_{m}=[L_{m}(t)]^{F_{n}}_{t=1}~~~~~~L_{m}(t)\in \Re
\end{equation}

Given joint location sequence $L_{m}$, first we use Tukey's test to detect any value in $L_{m}$ that is below $Qu_{low}-1.5\times IQR$, or above $Qu_{up}+1.5\times IQR$ where $IQR$ stands for the interquartile range, $Qu_{low}$ and $Qu_{up}$ are lower and upper quartile, respectively. If $o_{L_{m}}$ is the set of all the detected outlier indices in $L_{m}$ (each index corresponds with one time instant $t$) defined as 
\begin{equation}
\label{eq:outlierIndices}
\begin{cases}
       \text{$o_{L_{m}}=[o_{1},o_{2},...,o_{R}]$}\\
       \text{$o_{1} < o_{2} < ... < o_{R} $}\\
       \text{$o_{i}\in [1,2,...,F_{n}];i\in[1,2,...,R]$}\\
\end{cases}  
\end{equation}
where $R$ is the number of outliers in $L_{m}$ detected by the Tukey's method, then $L_{m}(o_{i})$ will be corrected according to the following 
\begin{equation}
\label{eq:OutlierCorrection}
 L^{c}_{m}(o_{i})=L^{NO}_{m}~~~~~~L^{NO}_{x}=1-NN(L_{m}(o_{i}))  
\end{equation}
where $L^{c}_{m}(o_{i})$ is the corrected value of $L_{m}$ at $t=o_{i}$. $NO$ and $1-NN$ stand for non-outlier, and the one nearest neighbor, respectively. $L^{NO}_{m}$ is the value of the nearest neighbor of $L_{m}(o_{i})$, that is not an outlier. In those cases with two nearest neighbors, one is selected randomly. After the detected outlier values of $L_{m}$ are corrected according to equation \ref{eq:OutlierCorrection}, piece-wise cubic Hermite polynomials \cite{fritsch1980monotone} are utilized to interpolate the missing values in $L_{m}$. We use piece-wise Hermite polynomial to preserve the shape of $L_{m}$. Meanwhile, by applying outlier correction before missing value interpolation, the shape of the curves will be less affected by outliers. Finally, we employ RLowess (locally weighted scattered plot smoothing) filter \cite{cleveland1979robust} to smooth the resulting joint location sequence and alleviate the effect of remaining smaller spikes in $L_{m}$. RLowess assigns a value to each point by locally fitting a first-order polynomial, utilizing weighted least squares. Weights are computed using the median absolute deviation (MAD), which is a robust measure of variability in the data in the presence of outliers. The robustness of weights is critical due to the existence of smaller-amplitude spikes that act as outliers. 
\begin{figure*}[!t]
\centering
\includegraphics[width=\textwidth]{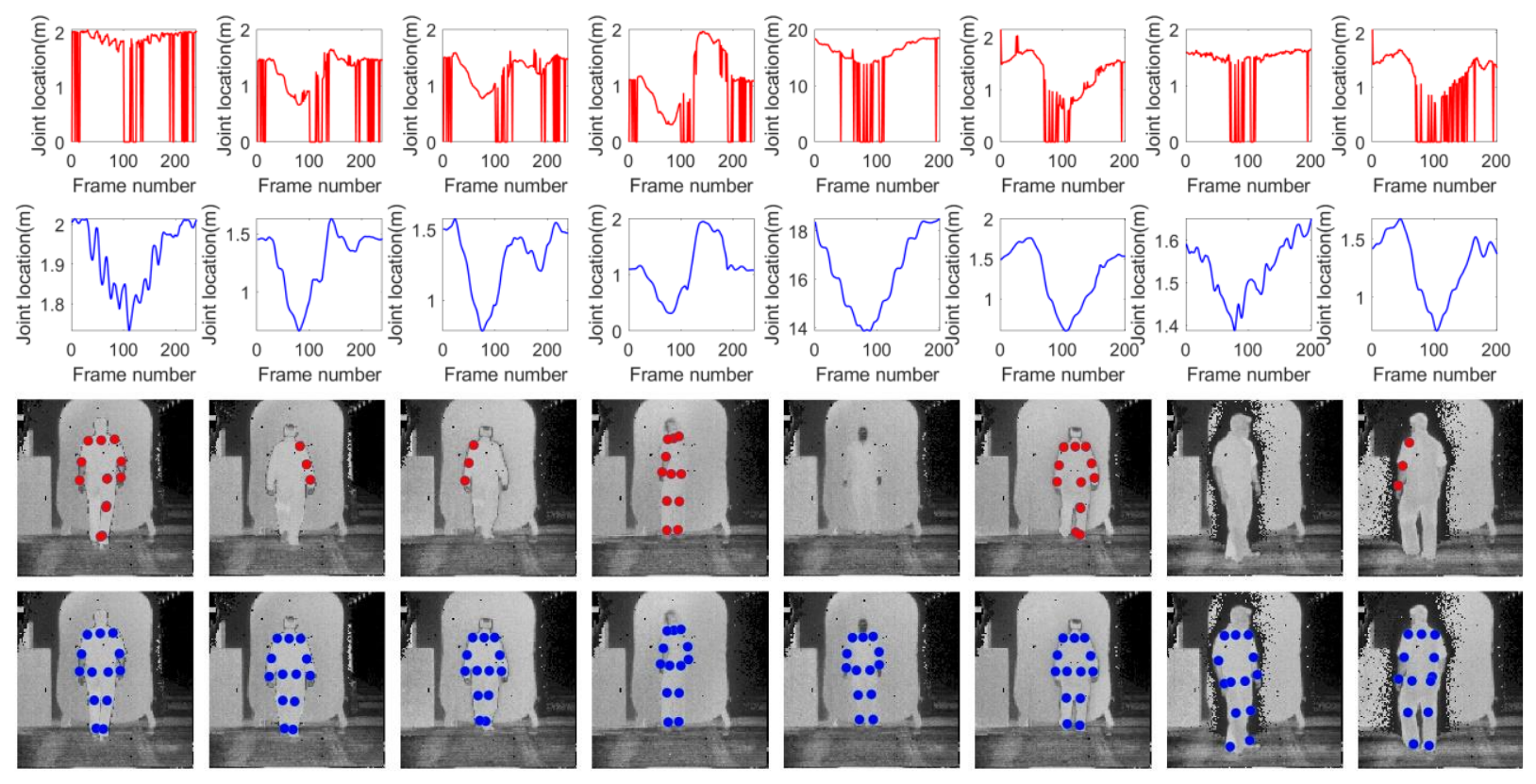}
\caption[Skeleton Joints]{Effect of joint location sequence filtering. From top: sample joint location sequences before (first row) and after (second row) joint location sequence filtering. Samples of faulty and missing skeleton joints before (third row) and after (bottom row) joint location sequence filtering.}
\label{fig:EffectOfJointFiltering}
\end{figure*}

\begin{figure}[!b]
\centering
\includegraphics[width=\linewidth]{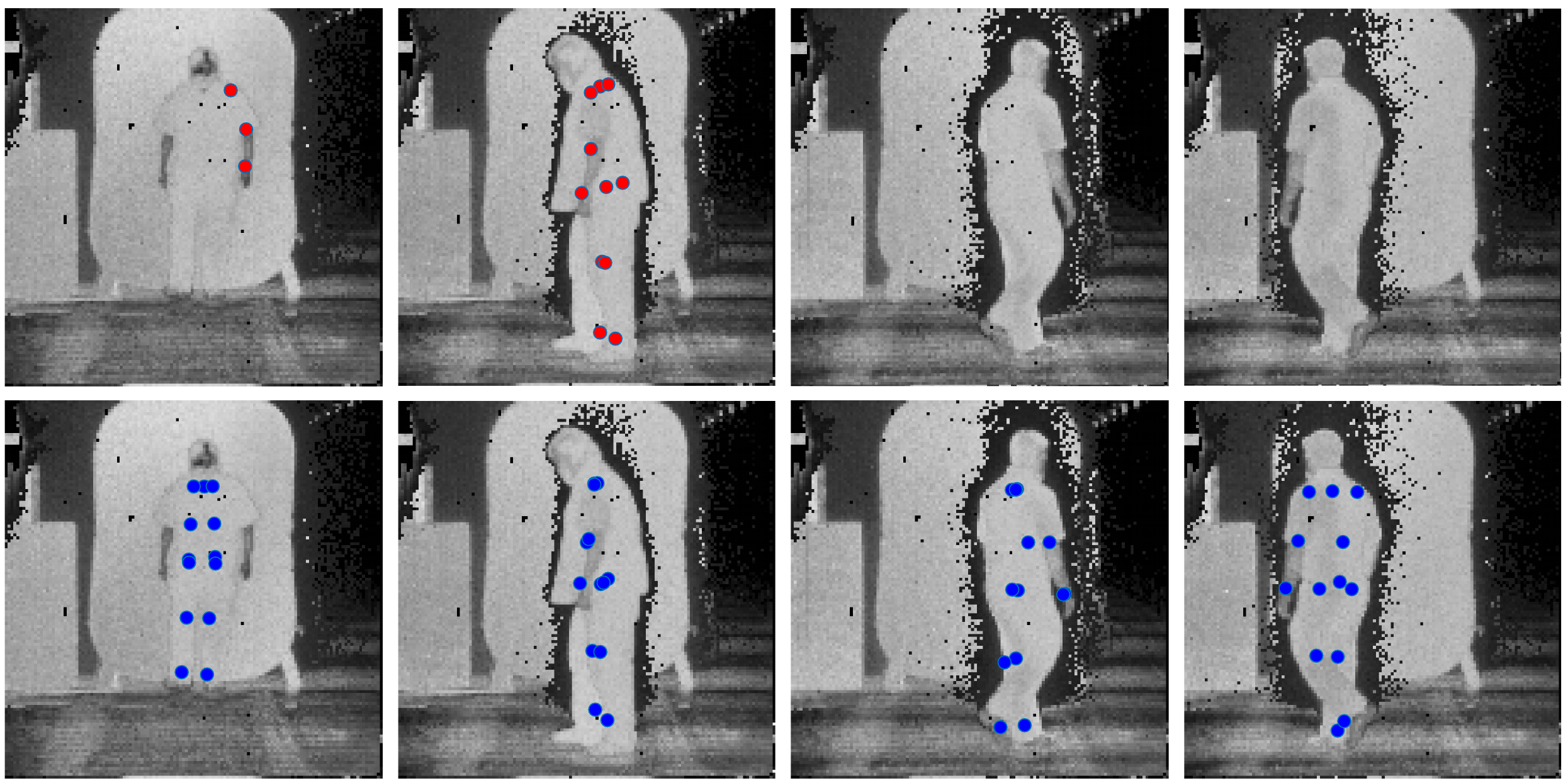}
\caption[Joint location failures]{Failure examples of joint sequence filtering. Sample frames of skeleton joints, before (top) and after (bottom) joint sequence filtering.}
    \label{fig:JointCorrectionFailures}
\end{figure}

The described filtering procedure will effectively correct joint location time sequences. Furthermore, when pose-detector fails to detect a skeleton model, the joint location filtering can interpolate the missing skeleton joint locations. Figure \ref{fig:EffectOfJointFiltering} illustrates the result of filtering on samples of joint location time sequences. As we can see in this figure, the original joint location sequences are noisy, containing many missing values and outliers. We can also see the results in the image reference frame, where missing joints are interpolated successfully through the filtering mechanism. While in the majority of cases, the interpolation of missing or noisy joints follows the correct joint locations, there exist cases where the obtained localization results are not accurate. Figure \ref{fig:JointCorrectionFailures} shows some failure examples in joint localization correction. However, even for failure cases, at least half of the joints are predicted correctly. This can enhance the likelihood of correct identification compared to the original localization of the joints.

\subsection{Incorporating the dynamics}
As humans, we recognize a familiar person not just by looking at their body measurements like height; we also incorporate the way that a person walks or moves their body in recognizing one subject from another subject. In the gait recognition language, the first set of features that are computed from body measurements like limb lengths or height are called static features. Attributes like step length or speed that comprise the motion of gait from one posture to another posture, are dynamic features. When individuals with approximately the same body measurements are considered, dynamic features are critical for a successful gait recognition. Speed, step length and stride length are among the widely used features to incorporate the dynamic of the motion \cite{preis2012gait, koide2016identification}. Another common practice in the majority of  model-based methods involves computing moments like mean, maximum, and variance of selected features over the length of each gait cycle \cite{sinha2013person,yang2016relative,chi2018gait}. The time sequence of the distance between the two ankle joints is a commonly employed attribute to compute the gait cycle. This practice has repeatedly proven to be successful in encoding the dynamic of the motion, achieving high accuracy in gait recognition. However, this analysis is commonly performed on a clean dataset that is recorded under controlled conditions, like limited directions of motion in front of the camera.
\begin{figure}[!t]
\centering
\includegraphics[width=\linewidth]{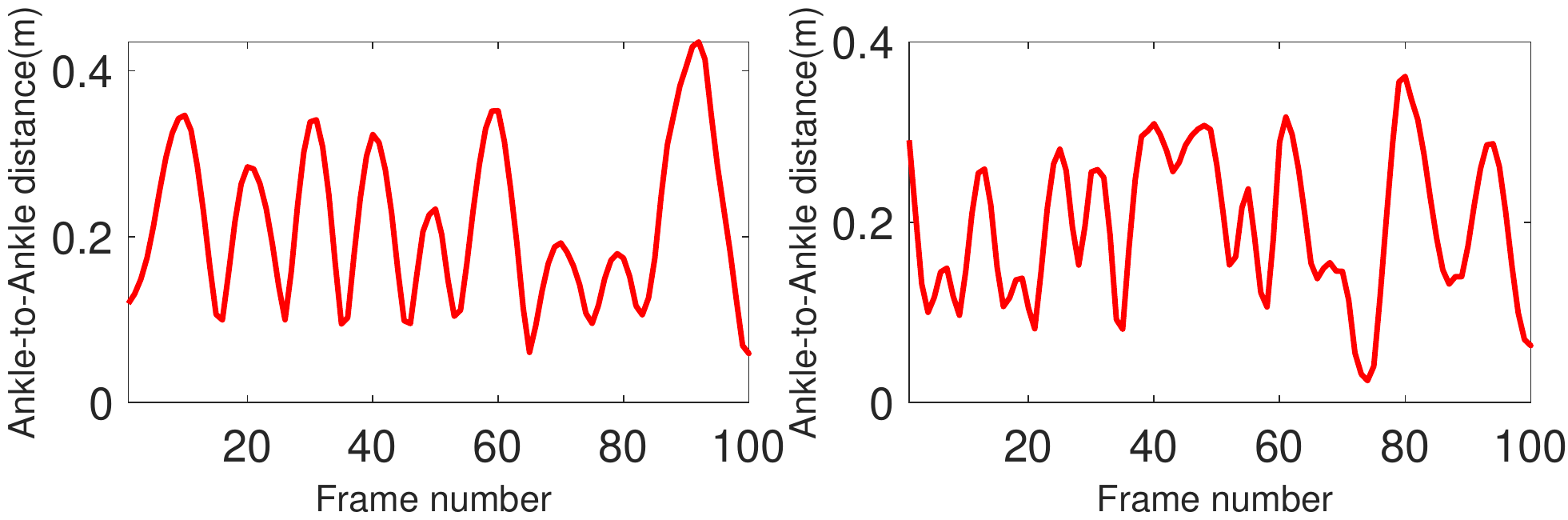}
\caption[lidar Ankle-to-Ankle]{Examples of time sequence of ankle to ankle distance of lidar data after joint correction. While the plot on the left presents a clear periodic pattern, the sequence on the right lacks such a pattern.}
\label{fig:CyclicAnkleToAnkleLidar}
\end{figure}

Figure \ref{fig:CyclicAnkleToAnkleLidar} shows examples of ankle to ankle distance time sequences for lidar data after joint location filtering. The sequence on the left shows a periodic pattern, however like the plot on the right side, there are many examples of such sequences that lack a clear cyclic pattern. In contrast with the lidar data, we generally observe a periodic pattern with the Kinect measurements. To resolve this issue, we incorporate statistics that are robust to noisy data. Joint sequence filtering improves the quality of gait features, and therefore as we will show later gait recognition accuracy. However, there is a considerable amount of consecutive frames with missing skeleton in each sequence. This will cause the result of joint sequence correction prone to noisy measurements. To compensate for this shortcoming, in addition to mean, standard deviation, and maximum, we include median, upper and lower quartiles that are robust to noisy data. This property is, in particular, beneficial for gait cycles that are corrupted with outlier features. We build feature vectors that comprise \textit{mean, standard deviation, maximum, median, lower quartile and upper quartile} of each feature over each gait cycle. Later, we will show that the resulting feature vectors can improve the classification scores over the feature vectors that only incorporate non-robust moments.

\subsection{Outlier removal}
\begin{figure}[!ht]
\centering
\includegraphics[width=\linewidth]{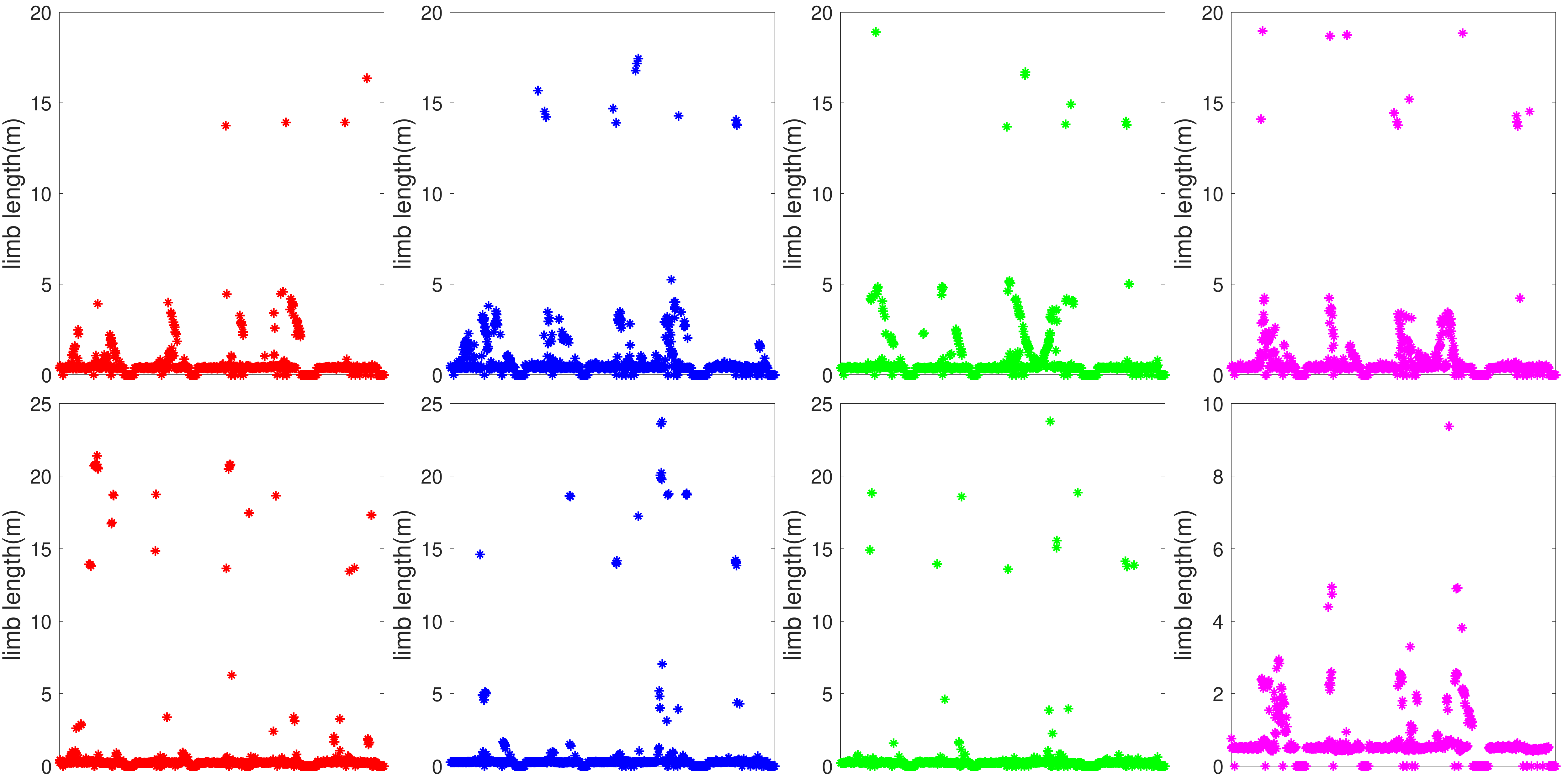}
\caption[Outlier Abundance]{Sample limb lengths for one subject from lidar data that shows abundance of outliers. Each graph represents the distribution of one limb length.}
\label{fig:outlierAbundance}
\end{figure}

Outliers are a set of observations that cannot be described by the underlying model of a process. While in some applications, i.e. surveillance and abnormal behavior detection, outlier observations can be of interest and are kept for further investigation, there are situations that outliers are the result of faulty measurements or caused by noise. The latter type of outliers have to be detected and removed before model estimation, because the models that are estimated utilizing the data which is contaminated by such outliers, are not accurate and generate many false predictions. For gait recognition, one common approach is to remove outlier measurements from the collected data by setting some measurement thresholds \cite{yang2016relative,liu2016spatio,semwal2017optimized,chi2018gait}. For comparison, and as an alternative approach to deal with the noisy and missing joint location measurements in our dataset that results into outlier features, we employed the Tukey method to detect outliers in the feature vectors that are computed from faulty and missing joint locations. The second row in Figure \ref{fig:FramesWithSkeleton} presents some of the examples of faulty skeletons that are the result of erroneous joint localization. Furthermore, there are frames with missing skeletons. Figure \ref{fig:outlierAbundance} shows selected limb lengths of one subject computed from joint coordinates extracted from flash lidar data. The joint data are not treated for correction and by looking at the scale and distribution of each limb length, we can clearly see the features are highly contaminated by outlier values. We use Tukey's test for outlier detection and employ it on every feature in a feature vector. We choose Tukey's test in particular to avoid making any assumption about the underlying distribution of the features.

\begin{figure}[!hb]
\centering
\includegraphics[width=\linewidth]{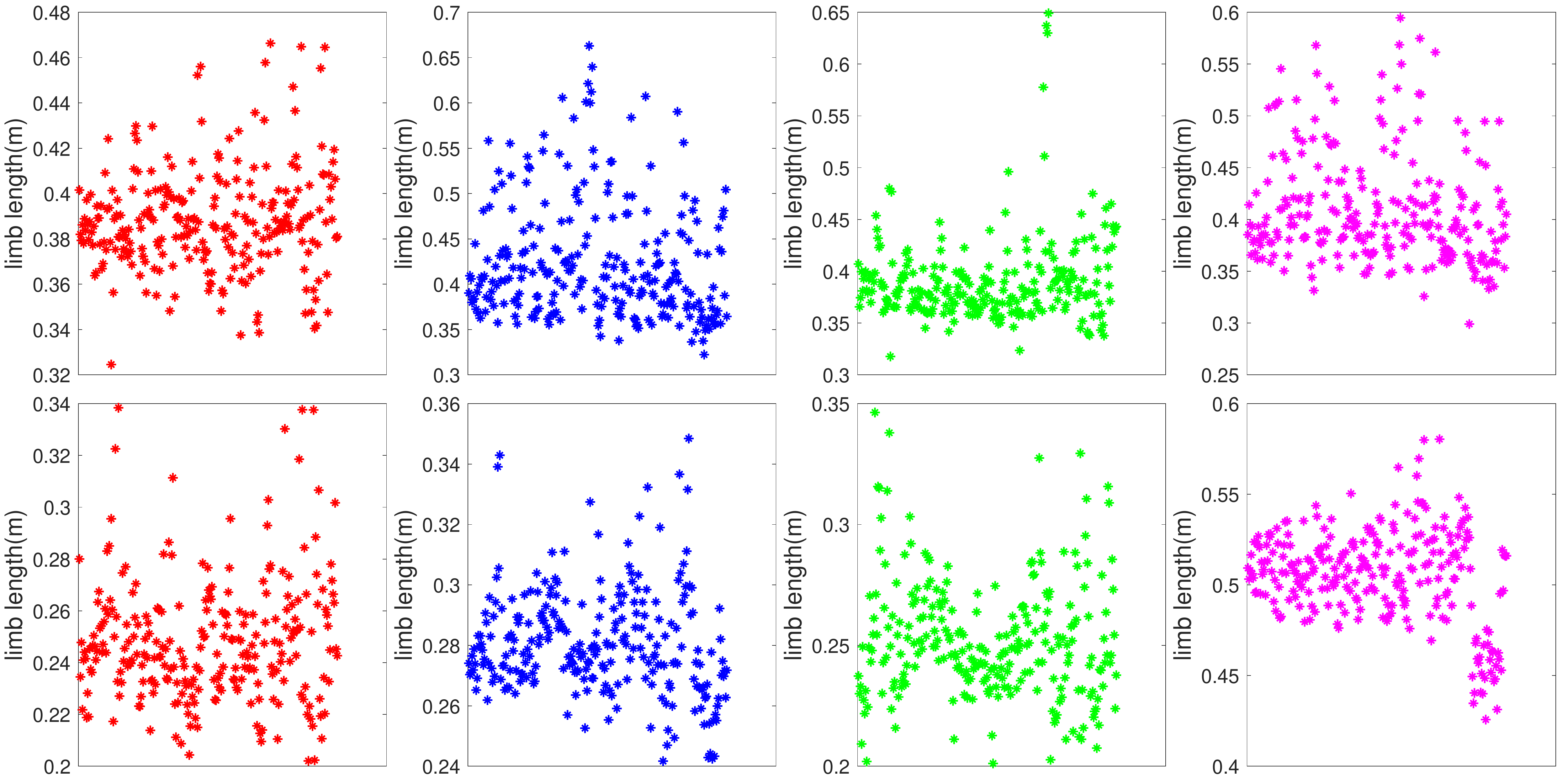}
\caption[After Tukey]{Same limb lengths as in Figure \ref{fig:outlierAbundance} after outlier removal. Compare the distribution and range of each limb length between the two figures.}
\label{fig:OutliersRemoved}
\end{figure}

We define $Jd=[Jd_{1}, Jd_{2}, ..., Jd_{P}]$ as a given feature vector, where $P$ is the number of features in $Jd$ and $Jd_{i}$ is the Euclidean distance between two skeleton joints. Before applying Tukey's test, first we remove all the frames with missing skeletons. Next, we filter the remaining features, by setting an upper threshold $T_{upper}$ that will be applied to all the features. To determine $T_{upper}$, we investigate the distribution of $Jd^{S}$
\begin{equation}
\label{eq:maxStdFeature}
Jd^{S}=max_{std}(Jd_{i})|^P_{i=1}
\end{equation}
where $Jd^{S}$ is the feature with maximum standard deviation. The histogram of $Jd^{S}$ is computed, and the maximum value of histogram bin interval $W$ is selected as $T_{upper}$ according to 
\begin{equation}
\label{eq:FilterThreshold}
\begin{cases}
       \text{$T_{upper}=max(W)$}\\
       \text{$Freq(W) \leq \alpha \times Freq(W_{max}) $}\\
       \text{$min(W)\geq max(W_{max})$}\\
\end{cases}   
\end{equation}
where $W_{max}$ is the histogram bin with the highest frequency, and $\alpha=.1$ is a hyperparameter, which is set according to the distribution of $Jd^{S}$. A feature vector with a feature that is beyond $T_{upper}$ will be removed. Next, Tukey's test is employed on each feature. $Jd$ is not an outlier if
\begin{equation}
\label{eq:TukeyOnDistanceFeatures}
Tukey(\{Jd_{i}\}^{P}_{i=1})=\mathbf{0}_{P}~~~~~where~~~~~Jd_{i}\in \Re^{+}
\end{equation}
where $\mathbf{0}_{P}$ is zero vector of length $P$. For feature $Jd{i}$, $Tukey(Jd_{i})=0$ means that $Jd_{i}$ passed the Tukey's test, or $Jd_{i}$ is not an outlier. Based on Equation \ref{eq:TukeyOnDistanceFeatures} for feature vector $Jd$ to be a non-outlier, all of its feature components have to be non-outliers. This means that $Jd$ is an outlier if there exists a $Jd_{i}$, such that $Tukey(Jd_{i})=1$. Figure \ref{fig:OutliersRemoved} presents the same features as in Figure \ref{fig:outlierAbundance} after outlier removal. By comparing the scale and values of features between the two figures, we observe a considerable reduction in the range of each feature as a result of outlier removal. This however, comes at the cost of eliminating a large portion of the data.

\subsection{Outlier removal for vector-based features} There are cases when the components of a feature vector are vectors. This happens if we compute the 3-dimensional vectors between skeleton joints. In other words, we have a $3\times Q$ vectorized matrix $Jv^{3D}=[Jv^{3D}_{1}, Jv^{3D}_{2}, ..., Jv^{3D}_{Q}]$ of the joint coordinates. $Q$ is the number of 3-dimensional vectors in $Jv^{3D}$, and $Jv^{3D}_{i}$ represents the $i-th$ column, which is the 3-dimensional vector between two skeleton joints 
\begin{equation}
\label{eq:Each3DVector}
Jv^{3D}_{i}=[x_{i},y_{i},z_{i}] \in \Re^{3N} 
\end{equation}
We need to treat each of the 3-dimensional vectors as one entity, rather than treating each dimension separately. In order to detect outliers for this set of features, we use the concept of marginal median. The marginal median of a set of vectors is a vector where each of its components is the median of all the vector components in that direction. We then use cosine distance to calculate vector similarity between each set of 3-dimensional vectors with their corresponding median vector. Defining $Jv^{median}$ as the marginal median over all given $Jv^{3D}$ feature vectors
\begin{equation}
\label{eq:CosineSimilarity}
S^{3D}=cos(Jv^{median}_{i},Jv^{3D}_{i})|^{Q}_{i=1} 
\end{equation}
where $S^{3D}_{i}=cos(Jv^{median}_{i},Jv^{3D}_{i})$ is the cosine similarity between i element of feature vector $Jv^{3D}$ and $Jv^{median}$. This procedure will create the cosine similarity measure between each $Jv^{3D}$ and the median vector $Jv^{median}$. Then Tukey's test is employed on the cosine similarity measures, and a feature vector is labeled as an outlier if at least one of its features is an outlier. Algorithm below describes outlier detection on the feature vectors built from 3-dimensional vectors using the concept of marginal median, cosine distance similarity measures between vectors, and Tukey's test.
\\ 
\hrule
\vspace{.2cm}
\noindent\textbf{Outlier detection for 39-D feature vectors}\\
\indent\textbf{1}.~Over all the given feature vectors, calculate the marginal \\
\indent ~~~median vector. Let the resulting median feature vector \\
\indent ~~~be $Jv^{median}$\\
\indent\textbf{2}.~For each 3D vector $Jv^{3D}_{i}$ in each feature vector $Jv^{3D}$, \\
\indent ~~~calculate $cos(Jv^{median}_{i},Jv^{3D}_{i})$; save the results in one\\ \indent ~~~row of $S$.\\
\indent\textbf{3}.~Employ Tukey's test on each row of $S$.\\
\indent\textbf{4}.~A given feature vector $Jv^{3D}$ will pass Tukey's test, if its\\ 
\indent ~~~ corresponding row in $S$ passes Tukey's test.\\
\vspace{.1cm}
\hrule
\vspace{.2cm}

\subsection{Feature vectors}
\begin{figure}[!b]
\centering
\includegraphics[width=.7\linewidth,height= 1.8in]{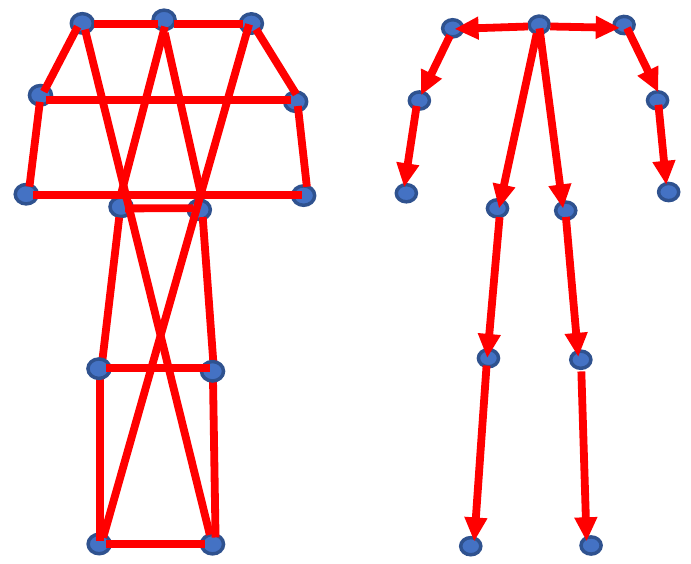}
\caption[Different Feature Vectors]{Illustration of two types of feature vectors: distance-based feature vector (left), vector-based feature vector (right). All The features are depicted in red color.}
\label{fig:DifferentFeatureVectors}
\end{figure}

To evaluate the performance of the proposed method, we use two different sets of feature vectors: length-based feature vectors and vector-based feature vectors. The length-based feature vector consists of a set of limb lengths and distance between selected joints in the skeleton that are not directly connected. This feature vector can be described similar to $Jd$ in \textit{"Outlier removal"} section, where $P=19$. Table \ref{tab:Length-based feature Vector} describes the components of the length-based feature vector. This set includes static limb length features and some other distance attributes that change during motion and encode information about postures. Figure \ref{fig:DifferentFeatureVectors}, left side presents an illustration of the length-based feature vector. 
\begin{table}[!ht]
	\centering
	\caption[length-based features]{List of length-based feature vectors (L refers to the left joints and R refers to the right joints)}
\label{tab:Length-based feature Vector}
\renewcommand{\arraystretch}{.85}
	\resizebox{.9\columnwidth}{!}{
	\begin{tabular}{p{4cm}p{4cm}}
		\toprule
		{\textbf{Feature}} & {\textbf{Feature}} \\ \midrule
		R and L Shoulder & Elbow to elbow  \\
		R and L upper arm & Wrist to wrist \\
		R and L lower arm & Hip to hip \\
		Spine & Knee to knee\\
		R and L upper leg & Ankle to ankle\\
		R and L lower leg & R shoulder to L ankle\\
		shoulder to shoulder & L shoulder to R ankle\\ \bottomrule
	\end{tabular}
	}
\end{table}

The second set of feature vectors is vector based. This means that each feature is a 3-dimensional vector, computed between two skeleton joints. Compared to distance-based features \cite{yang2016relative}, or to the angle-based attributes \cite{ball2012unsupervised}, vector-based features encode the angle and distance between selected joints of the skeleton. Table \ref{tab:featureVector} lists the joints that form each of the 3-dimensional vectors in the vector-based feature vector. This feature vector can be described similar to $Jv^{3D}$ in the last section, where $Q=12$. Unlike features in \cite{kumar2012human} that are computed with respect to a reference joint, the vectors in the vector-based feature vector are formulated between different joints, mimicking the limb vectors in the skeleton model. An illustration of the vector-based features is given in the right side of Figure \ref{fig:DifferentFeatureVectors}.
\begin{table}[!b]
\small
	\centering
	\caption[vector-based features]{List of three-dimensional vectors in the feature vector (L refers to the left joints and R refers to the right joints)}
\label{tab:featureVector}
\renewcommand{\arraystretch}{.95}
	\resizebox{.9\columnwidth}{!}{
	\begin{tabular}{p{4cm}p{4cm}}
		\toprule
		{\textbf{3D vector}} & {\textbf{3D vector}} \\ \midrule
		Neck to R Shoulder & R Hip to R Knee \\
		Neck to L Shoulder & L Hip to L Knee \\
		Neck to R Hip & R Elbow to R Wrist \\
		Neck to L Hip & L Elbow to L Wrist \\
		R Shoulder to R Elbow & R Knee to R Ankle\\
		L Shoulder to L Elbow & L Knee to L Ankle\\ \bottomrule
	\end{tabular}
	}
\end{table}

\section{Results and discussion}
\subsection{TigerCub 3D Flash lidar} The TigerCub is a light-weight 3D flash lidar camera that provides real-time range and intensity data, using eye-safe Zephyr laser \cite{horaud2016overview}. The performance of the camera is not affected by the lack of light at night, or in the fog or dust. Like other lidar cameras, it can provide a detailed 3D mapping of the scene, where close objects can be recognized from each other. These properties make flash lidar a suitable candidate for real time data acquisition and autonomous operations.

TigerCub 3D Flash lidar has a focal plane of $128\times 128$, and can stream up to 20 frames per second.  
\subsection{Dataset} The dataset in this work has been recorded using a single TigerCub 3D Flash lidar camera, where the camera is located in a fixed location during all the actions. There are in total 34 sequences of walking actions performed by 10 subjects. The recording includes walking action of three main categories; walking toward and away from the camera, walking on a diamond shape, and walking on a diamond shape while holding a yard stick with one hand. For those frames in which subjects walk toward and away from the camera, all the views are from the front and back of the person, plus some frames of side views when the subjects turn away. The sequences with walking on a diamond shape offer more frames with the side views of the subjects. The data is captured at the rate of $15$ fps with $128\times 128$ frame resolution. The number of frames per video is different, with $130$ frames for the shortest video to $498$ frames for the video with the highest number of frames. Table \ref{table:VideoFrames} shows the number of frames per subject for each category of the walking action. Each frame has two sets of data, intensity and range, both with the same number of pixels, where intensity data is in gray-scale and the range data shows the distance of each point in the field of view from the camera sensor. 
\begin{table}[!t]
	\centering
\caption[Video Frames]{Number of frames per type of walking action for each subject. FB Walk: front back walk, D Walk: diamond walk, DS Walk: diamond walk holding stick}
\label{table:VideoFrames}
	\resizebox{.9\columnwidth}{!}{
	\begin{tabular}{@{}ccccc@{}}
		\toprule
		{} & {\textbf{FB Walk}} & {\textbf{D Walk}} & {\textbf{DS Walk}} & {\textbf{Total}} \\ \midrule
		subject 1 & 130 & 215 & 463 & 808\\
		subject 2 & 248 & 462 & 451 & 1161\\
		subject 3 & 199 & 398 & 391 & 988 \\
		subject 4 & 224 & 377 & 405 & 1006 \\
		subject 5 & 257 & 459 & 486 & 1202 \\
		subject 6 & 226 & 483 & 881 & 1590 \\
		subject 7 & 204 & 429 & 394 & 1027 \\
		subject 8 & 249 & 474 & 445 & 1168 \\
		subject 9 & 203 & 897 & 375 & 1475 \\
		subject 10 & 216 & 441 & 385 &  1042\\ \bottomrule
	\end{tabular}
	}
\end{table}

\begin{table}[ht]
	\centering
	\caption[results before correction]{Correct identification scores (average accuracy and F-score) for the proposed features and the other methods. LB stands for length-based feature vector, and VB stands for vector-based feature vector. Features are computed without joint correction.}
	\label{table:ResultsComparisonOriginalData}
	\resizebox{.9\columnwidth}{!}{
	\begin{tabular}{@{}ccc@{}}
		\toprule
		{\textbf{Method}} & {\textbf{Average Accuracy(\%)}} & {\textbf{Average F-score(\%)}} \\ \midrule
		\cite{preis2012gait}  & 27.90    & 25.36  \\
		\cite{ball2012unsupervised}   & 25.34 & 23.24  \\
		\cite{sinha2013person}   & 61.81  & 54.61 \\
		\cite{yang2016relative}  & 63.82  & 58.64 \\
		GlidarCo, LB     & 54.96 & 51.58\\ 
		GlidarCo, VB     & \underline{67.16} & \underline{63.47}\\ \bottomrule
	\end{tabular}
	}
\end{table}
\subsection{Performance comparison} To evaluate the performance of the purposed method, we carry out a comparison with four state-of-the-art relevant gait recognition methods, the work of Preis \cite{preis2012gait}, Ball \cite{ball2012unsupervised}, Sinha \cite{sinha2013person}, and Yang \cite{yang2016relative}. Preis \textit{et al}. use a set of static features, plus step length and speed as dynamic features. In \cite{ball2012unsupervised}, authors use the moments of six lower body angles. Sinha combines the features in \cite{preis2012gait} and \cite{ball2012unsupervised} with their own area-based and distance between body segments features. Yang \textit{et al}. utilize selected relative distance along different motion direction. The performance comparison includes the average accuracy and F-score as a measure of effectiveness of each method. In our experiments, we also consider the outlier removal method as another alternative approach and compare its performance with the other methods. Furthermore, to investigate the effectiveness of joint correction filtering, we compare the performance of all the methods after joint correction. We also evaluate joint correction effect on length-based and vector-based feature vectors. We use $75\%$ of the sequences for training and the rest for testing. To insure the generalization of the proposed method, the classifier is tested on a type of walking that it was not trained on. Support vector machine (SVM) with the radial basis function (RBF) kernel is adopted as our classifier. Our vector-based and length-based features are computed per frame and no over-the-cylce moment computation is performed. Therefore, in this experiment, we do not incorporate motion dynamics in our features. 

\begin{table}[h]
\huge
\centering
	\caption[results after correction]{Correct identification scores (average accuracy and F-score) for the proposed features and the other methods. Features are computed from the joint locations  corrected by the proposed joint correction filtering.}
\label{table:ResultsComparisonCorrectedData}
	\resizebox{.9\columnwidth}{!}{
	\begin{tabular}{@{}ccc@{}}
		\toprule
		{\textbf{Method}} & {\textbf{Average Accuracy(\%)}} & {\textbf{Average F-score(\%)}} \\ \midrule
		\cite{preis2012gait}  & 41.07  & 38.59  \\
		\cite{ball2012unsupervised}   & 28.33 & 26.25  \\
		\cite{sinha2013person}   & 80.84  & 78.96 \\
		\cite{yang2016relative}  & 75.19  & 70.50 \\
		Outlier removal, LB & 76.60 & 68.89 \\
		Outlier removal, VB & 80.70 & 75.22 \\
		GlidarCo, LB & 76.37 & 70.19 \\ 
		GlidarCo, VB     & \underline{84.88} & \underline{78.98}\\ \bottomrule
	\end{tabular}
	}
\end{table}

Table \ref{table:ResultsComparisonOriginalData} shows the correct identification scores without joint correction. As we can see, the identification scores are generally low when features are computed from the skeleton data without correction filtering. This illustrates the fact that joint location coordinate values are noisy, therefore the resulting erroneous features jeopardize a successful gait identification. Results in Table \ref{table:ResultsComparisonCorrectedData} report identification scores with the proposed joint correction. It also shows the scores when outlier removal is applied on the features. While outlier removal can improve the identification scores, it is not as effective as joint correction. This might be caused by the noisy features that still exist after outlier removal, which can be observed by looking at the range of selected limb lengths after outlier removal in Figure \ref{fig:OutliersRemoved}. Furthermore, outlier removal results into elimination of more than $40\%$ of the data, which can be problematic when data is limited. The results in Table \ref{table:ResultsComparisonCorrectedData} demonstrate the effectiveness of joint correction, where it improves the gait identification scores in all of the cases. Among the evaluated methods, the performance of \cite{ball2012unsupervised} does not improve as much as the other approaches. In \cite{ball2012unsupervised} authors use six angles between lower body joints as the features and compute three moments of each angle over every gait cycle. We see in Figure \ref{fig:JointAndSkeleton} that the adopted skeleton model in our work lacks the foot joints that are essential to estimate two of the angles in \cite{ball2012unsupervised}. To calculate these angles, we estimate the floor plane and use the normal vector to the plane. We speculate the error in this estimation might also incorporate into lower performance of this method compared to the others. Furthermore, it was reported before that distance-based features might work better than angle-based features, in particular when the number of subjects is relatively low \cite{dikovski2014evaluation}. Joint angles are also prone to changes in the walking speed \cite{han2015influence,kovavc2014human}. We also observe that regardless of the feature type, both length-based and vector-based features perform better after joint correction filtering. By comparing the results in both Table \ref{table:ResultsComparisonOriginalData} and \ref{table:ResultsComparisonCorrectedData}, we also realize that vector-based features outperform length-based features. Furthermore, while our features do not contain the dynamics of the motion, vector-based features still outperform methods that incorporate temporal information by computing moments of features over gait cycle.

\begin{table}[!t]
	\centering
	\caption[Moments over gait cycle]{Correct identification scores (average accuracy and F-score) with statistics of features computed over gait cycle. LB refers to length-based features, and VB refers to the vector-based features. the 3 statistics case refers to computing only mean, maximum, and standard deviation of each feature over every gait cycle. 6 statistics scenario adds median, lower and upper quartile to the initial 3 statistics.}
	\label{table:MomentsOverGaitCycle}
	\resizebox{.9\columnwidth}{!}{
	\begin{tabular}{@{}ccc@{}}
		\toprule
		{\textbf{Method}} & {\textbf{Average Accuracy(\%)}} & {\textbf{Average F-score(\%)}} \\ \midrule
		LB (3 statistics) & 70.50 & 66.75  \\
		LB (6 statistics) & 75.22 & 73.22 \\
		VB (3 statistics) & 76.28 & 74.01 \\
		VB (6 statistics) &  \underline{84.65} & \underline{80.38} \\ \bottomrule
	\end{tabular}
	}
\end{table}

\subsection{Evaluating features over gait cycle}As we discussed earlier, we also compute six statistics of our features over each gait cycle to incorporate the motion dynamics. Table \ref{table:MomentsOverGaitCycle} presents the identification scores when the statistics of length-based and vector-based features are computed over each gait cycle. By comparing the classification scores, we make an interesting observation that adding median, upper, and lower quartile to mean, maximum, and standard deviation, which are the common statistics widely employed in many model-based methods, can improve the identification results. By comparing the results in Tables \ref{table:ResultsComparisonCorrectedData} and \ref{table:MomentsOverGaitCycle}, we see that identification accuracy using the statistics of features over each gait cycle (table (\ref{table:MomentsOverGaitCycle}) is almost the same as the per-frame method (table \ref{table:ResultsComparisonCorrectedData}). However, the F-score improves with the former method. The average per-class accuracy and F-score for the per-frame method is summarized in Table \ref{table:PerClassScoresPerFrame}. We also present the per-class accuracy and F-score for the gait cycle statistics in Table \ref{table:PerClassScoresGaitFeatures}. By comparing the per-class classification scores for the per-frame and statistics over gait cycle, we also see that the minimum per-class accuracy and F-score are improved by $11\%$ and $2\%$ as a result of employing gait cycle statistics. This implies that by including the motion dynamics through the feature statistics, we can improve the performance of our model in general. This also indicates that by employing features that encode the motion dynamics, we can build a more reliable model compared to features that only include the static features. Last, the results in Tables \ref{table:ResultsComparisonCorrectedData}, \ref{table:MomentsOverGaitCycle}, \ref{table:PerClassScoresPerFrame}, and \ref{table:PerClassScoresGaitFeatures} suggest that as we increase the number of subjects for the identification task, the gait statistics that include static features through a dynamic criterion become superior to the per-frame case, where only static attributes are considered. 
\begin{table}[!ht]
	\centering
	\caption[Per-class scores Per-frame]{Correct identification scores (average accuracy and F-score) for each class of subject for the per-frame scenario of vector-based features. The minimum, and the next-to-lowest accuracy and F-score are presented in underlined type.}
\label{table:PerClassScoresPerFrame}
	\resizebox{.9\columnwidth}{!}{
	\begin{tabular}{@{}ccc@{}}
		\toprule
		{\textbf{Method}} & {\textbf{Average Accuracy(\%)}} & {\textbf{Average F-score(\%)}} \\ \midrule
		subject 1 & 93.08 & 92.02  \\
		subject 2 & 91.54 & 73.46 \\
		subject 3 & 73.08 &  \underline{64.63}\\
		subject 4 & 83.85 & \underline{61.76} \\
		subject 5 & 96.15 & 84.75 \\
		subject 6 &  \underline{67.69} &  69.02 \\
		subject 7 & 100 & 84.69 \\
		subject 8 & 75.77 & 82.95 \\
		subject 9 & \underline{51.79} &  67.90\\
		subject 10 & 81.92 & 88.94 \\ \bottomrule
	\end{tabular}
	}
\end{table}

\begin{table}[!ht]
	\centering
	\caption[Per-class scores gait cycle]{Correct identification scores (average accuracy and F-score) for each class of subject for the statistics of vector-based features over gait cycle. The minimum, and the next-to-lowest accuracy and F-score are presented in underlined type.}
\label{table:PerClassScoresGaitFeatures}
	\resizebox{.9\columnwidth}{!}{
	\begin{tabular}{@{}ccc@{}}
		\toprule
		{\textbf{Method}} & {\textbf{Average Accuracy(\%)}} & {\textbf{Average F-score(\%)}} \\ \midrule
		subject 1 & 87.50 & 93.33 \\
		subject 2 & 75 & \underline{63.16} \\
		subject 3 & 75 & 75 \\
		subject 4 & 75 &  70.59\\
		subject 5 & 100 & 88.89 \\
		subject 6 & 100 &  \underline{69.57} \\
		subject 7 & 87.50 & 82.35 \\
		subject 8 & 87.50 & 90.32 \\
		subject 9 &  \underline{70.83} & 80.95 \\
		subject 10 & \underline{62.50} & 76.92 \\ \bottomrule
	\end{tabular}
	}
\end{table}

\subsection{Effect of the number of training samples} In real world scenarios, there is always the issue of limited data for the task of gait recognition. Therefore, it is essential to investigate how the designed model or the selected features perform under limited data availability. We study the effect of the number of training examples on the performance of the corrected data with the assigned feature vectors. For this experiment, we examine the effect of the number of training samples on the performance of the vector-based features, both for the per-frame approach as well as the statistics over gait cycle scenario.

\begin{figure}[!t]
\centering
\includegraphics[width=\linewidth]{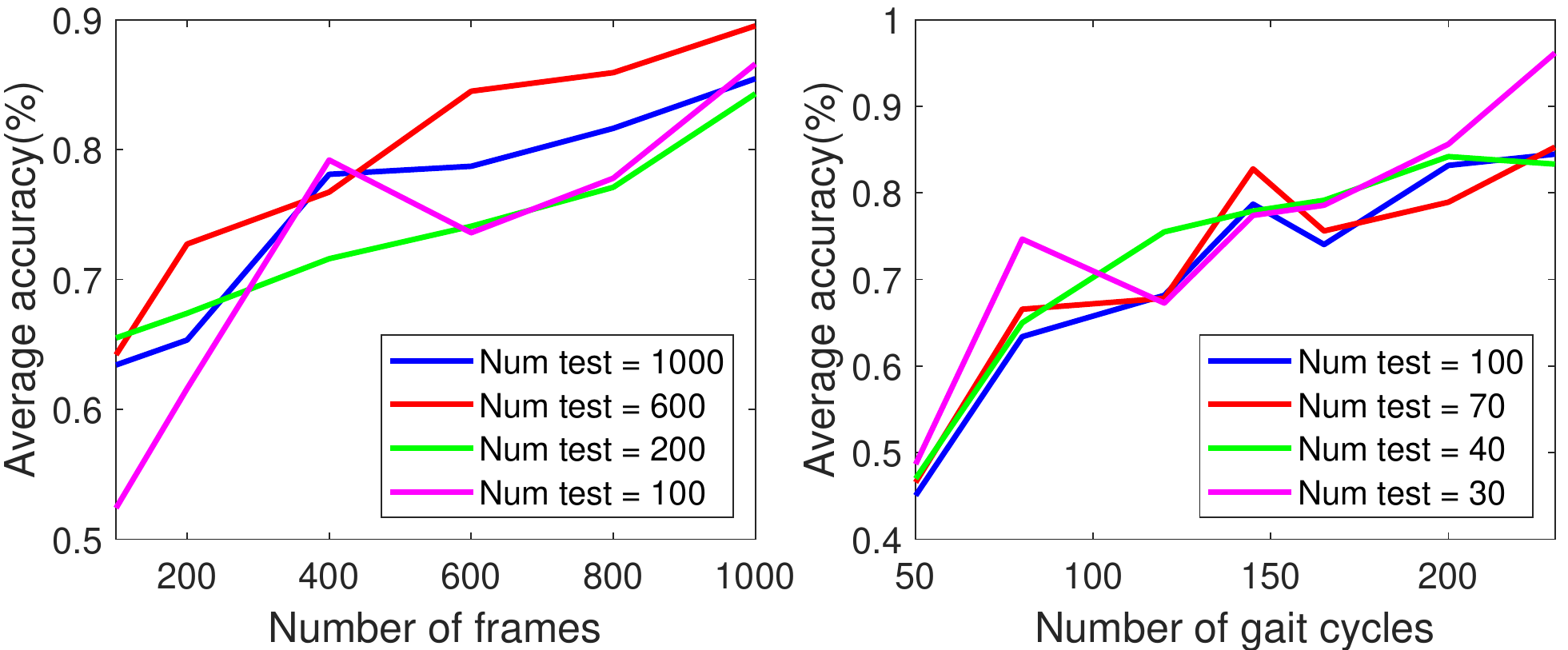}
\caption[Effect of training samples]{Average classification accuracy for different sizes of training sample sets given a number of test examples for the frame-based (left), and statistics over gait cycle-based (right).}
\label{fig:TrainingSamplesEffect}
\end{figure}

Figure \ref{fig:TrainingSamplesEffect}, left presents the identification accuracy as a function of the number of training examples, for several number of test feature vectors in the $[100,1000]$ range. For a given number of test samples, as we increase the number of training data, the accuracy of identification improves. When the size of test samples is small, accuracy increases at a higher rate as a result of using a larger number of training samples. A test sample size equal to or larger than 200 frames appears to be  a proper choice empirically, as the accuracy trend shows to be more stable. We also observe that the best performance is obtained with a training set of 1000 samples, irrespective of the number of test data. 

Figure \ref{fig:TrainingSamplesEffect}, right illustrates the same experiment for the number of gait cycles, when the statistics of features over gait cycle are considered as the feature vectors. The number of training cycles changes over the range of $[50,230]$, and the average classification accuracy is computed when different number of gait cycles is employed for testing. A comparison between the four different graphs in this figure illustrates that regardless of the number of test samples, using a training sample of at least $200$ gait cycles, we can acquire the highest classification accuracy with this feature vector. It should be noticed that while for test number $=30$ we can achieve a higher accuracy for training samples of size 200 and higher, this only occurs due to a limited number of test examples.  

\section{Conclusion} In this work, we presented a model-based gait recognition method using data collected by a flash lidar camera. The dataset contains 10 subjects, walking in three different manners in different directions. The detected skeletons from the collected sequences contain a considerable number of erroneous joint location measurements. Furthermore, the whole or part of the skeleton joints are missing in many frames. To improve the quality of the joint localization and to enhance gait recognition accuracy, we present GlidarCo. Unlike the common practice of removing noisy data under the described scenario, GlidarCo takes an unorthodox approach, by way of a filtering mechanism that corrects faulty skeleton joint positions to effectively improve the quality of joint localization and gait recognition. We also proposed a new and effective set of vector-based features that encode both length and angle of the limbs. Through the correction mechanism and the proposed vector-based features, GlidarCo obtained higher classification scores compared to state-of-the-art methods. Furthermore, to incorporate motion dynamics, robust statistics are integrated that can effectively improve the performance of the designed features that only employ traditional feature moments over the gait cycles. Future work will focus on anomaly detection in gait studies using lidar. 

%



\section*{Acknowledgment}

This work is funded in part by the U.S. Army DEVCOM, C5ISR  Center NVESD.

\ifCLASSOPTIONcaptionsoff
  \newpage
\fi



%


\bibliographystyle{IEEEtran}
\bibliography{bare_jrnl}


%








\end{document}